\documentclass[article]{IEEEtran}
\usepackage{algorithm}
\usepackage{algorithmic}
\usepackage{amsfonts}
\usepackage{float}
\usepackage{xcolor}
\usepackage{amsmath}
\usepackage{graphicx}
\makeatletter
\newcommand*{\da@rightarrow}{\mathchar"0\hexnumber@\symAMSa 4B }
\newcommand*{\da@leftarrow}{\mathchar"0\hexnumber@\symAMSa 4C }
\newcommand*{\xdashrightarrow}[2][]{%
  \mathrel{%
    \mathpalette{\da@xarrow{#1}{#2}{}\da@rightarrow{\,}{}}{}%
  }%
}
\newcommand{\xdashleftarrow}[2][]{%
  \mathrel{%
    \mathpalette{\da@xarrow{#1}{#2}\da@leftarrow{}{}{\,}}{}%
  }%
}
\newcommand*{\da@xarrow}[7]{%
  \sbox0{$\ifx#7\scriptstyle\scriptscriptstyle\else\scriptstyle\fi#5#1#6\m@th$}%
  \sbox2{$\ifx#7\scriptstyle\scriptscriptstyle\else\scriptstyle\fi#5#2#6\m@th$}%
  \sbox4{$#7\dabar@\m@th$}%
  \dimen@=\wd0 %
  \ifdim\wd2 >\dimen@
    \dimen@=\wd2 %
  \fi
  \count@=2 %
  \def\da@bars{\dabar@\dabar@}%
  \@whiledim\count@\wd4<\dimen@\do{%
    \advance\count@\@ne
    \expandafter\def\expandafter\da@bars\expandafter{%
      \da@bars
      \dabar@ 
    }%
  }%
  \mathrel{#3}%
  \mathrel{%
    \mathop{\da@bars}\limits
    \ifx\\#1\\%
    \else
      _{\copy0}%
    \fi
    \ifx\\#2\\%
    \else
      ^{\copy2}%
    \fi
  }%
  \mathrel{#4}%
}
\makeatother
\begin{document}
\title{Customer Empowered Privacy-Preserving Secure Verification using Decentralized
Identifier and Verifiable Credentials For Product Delivery Using Robots}
\author{Chintan Patel,~\IEEEmembership{Member,~IEEE}
\thanks{Chintan Patel is a Postdoctoral Researcher with the Department of Computer Science, University of Sheffield, UK.\protect\\
E-mail: chintan.p592@gmail.com}}
\maketitle
\begin{abstract}
In the age of respiratory illnesses like COVID-19, we understand the necessity for a robot-based delivery system to ensure safe and contact-free courier delivery. A blockchain-based Dynamic IDentifier (DID) gives people total power over their identities while preserving auditability and anonymity. A human mobile phone and a robot are machines created with a chip, making it simple to deploy a physical unclonable function-based verification system between the robot and the customer. This article presents a novel framework and a first customer verification scheme for verified courier delivery utilizing the blockchain-enabled DID and PUF-enabled robots. We employ DID for customer authentication between a robot (a service provider) and a customer and PUF for robot verification by the customer. We've also put the proposed work into practice and demonstrated its capabilities in terms of throughput, latency, computing cost, and communication cost. We also show formal security proof for the proposed user verification scheme based on the tamarin prover.
\end{abstract}
\IEEEdisplaynontitleabstractindextext
\IEEEpeerreviewmaketitle
\ifCLASSOPTIONcompsoc
\IEEEraisesectionheading{\section{INTRODUCTION}\label{sec:introduction}}
\else
\section{\textbf{INTRODUCTION}}\label{sec:introduction}
\fi
\noindent {\textbf{T}}{}he Robot-based product delivery is the future of the home delivery system \cite{Wang2022}. Nowadays, many big companies such as Amazon and kiwibot started to deliver a product using the robot. Amazon uses a six-wheeled robot named \textbf{Amazon Scout} to deliver a product. These all the robots are electric and help a lot in the UN's movement toward net-zero carbon emission by 2050. The robot-based delivery also achieves contactless product delivery that can help the world recover from pandemics such as COVID-19. 

For successful product delivery using a robot, verification of robot by customer and verification of customer by robot is a need of today. Most of the existing robot-based delivery companies use biometrics as a security verification factor or PIN as a verification factor. In a biometric-based verification system, a customer can not assign a product delivery to their friend or neighbour, while in the PIN-based delivery system, if the customer forgets a PIN or doesn't receive OTP on time, then it can lead to delivery failure. 

Protecting customer privacy is also a big challenge for e-commerce companies. Any Man-In-The-Middle (MITM) can sometimes trace the customer purchase pattern and customer purchase details. In 2021, 14 million amazon and eBay customer purchase data were leaked, which is a big concern for customer privacy. The only solution for this privacy and tracing problem is self-sovereign identity (SSI) and the use of a Decentralized Identifier (DID) as an identity of the customer that is generated new whenever the customer orders a product and is securely verifiable through the centralized registry such as blockchain. 

In this paper, we propose a novel (to the best of our knowledge first) solution with a novel DID-based framework for customer verification. We use DID as an identifier for customer identification and Physically Unclonable Function (PUF) for robot verification. Section 2 discusses the background and prelude related to the proposed work, followed by contributed work in section 3. Section 4 provides the security verification for the proposed work using the tamarin prover and key security properties. Section 5 discusses the experimental study with a key performance analysis matrix, followed by a conclusion and open challenges in section 6. 

\subsection{Related Work}

During the COVID-19 pandemic, contactless delivery has proven to be one of the most dependable methods for enabling social isolation and protecting sensitive groups. So, there are two possible approaches for this: The first one is drone-based delivery, and the second one is robot-based delivery. In this paper, we focus on the robot-based delivery of the product, and the proposed work can also apply to drone-based product delivery. In 2021, starship achieved 1 million autonomous deliveries using the robot. The tiny mile started to deliver products using robots in the Toronto area in 2019. In 2017, kiwibot achieved more than one lakh robot-based deliveries. So many other companies focus on robot-based delivery systems and future product delivery. Now a days, e-commerce customer uses same identity and password based mechanism that leads to several critical and fundamental security problems such as tracebility, password theft, wasting time if password loss and so many others. There is a \textbf{\textit{strong need to design a user empowered system}} for e-commerce industry where for each new order, customer generates new identity by self and get validated.   

The first paper directly related to securing a robot-based delivery system was proposed in 2021 by Wang et al. In this paper author uses artificial intelligence for authenticating the user. In 2022, Yang et al. proposed a customer authentication scheme using QR codes and asymmetric encryption. The author focuses on customer verification using biometric, QR code and PIN code technologies in the above-proposed work. Authors in \cite{Wang2022} don't focus on the security of complete e-commerce activity where the customer orders a product and the customer receives a product and it lacks from securing significant operations of product delivery. They use the same id each time for customer verification, leading to traceability and privacy problems also.  

In 2019, Kortesniemi et al. \cite{kortesniemi2019improving} presented an idea for improving privacy in the IoT ecosystem using DID. In 2020, Fedrecheski \cite{fedrecheski2020self} presented benefits and challenges related to self-sovereign identities for IoT environment and highlighted that DID resolution is a key challenge for constraint devices. In 2021, Samir et al. \cite{samir2021dt} presented a DID-based identity management framework with an authentication and integrity validation system for externally stored identities without the need for disclosing information related to identities. In 2022, Rohini et al. \cite{GopeDID2022} presented the first key exchange protocol for EV charging based on DID and claimed that the proposed protocol achieves privacy, accountability, anonymity and authentication. Even though, there are several proposal for DID based authentication, To the best of our knowledge, this is the \textbf{\textit{first DID-based customer verification scheme}} with physically unclonable function enabled robot-based product delivery in e-commerce services.  

\subsection{Motivation} 
In COVID - 19, people realize the need for a contactless delivery system, and robots can play a pivotal role in this work with their coequals, drones. The robot-based delivery system should ensure verification of both customers and the robot so that the robot can get assurance that the receiving person is the same who has ordered a product. At the same time, the customer must ensure that the robot from whom they are receiving a courier is a robot of the same company on which they have placed the order. Nowadays, most e-commerce companies use a centralized approach where customers must register their identities with trusted servers. There is a high probability that customers may lose their privacy if outsiders or insiders compromise this trusted server. To the best of our knowledge, this is the first work focusing on robot verification based on PUF. A PUF ensures unique randomness during verification through challenge-response pairs, leading to secured verification of chip-based machines. In the proposed work, DID stored in a permissioned blockchain solves numerous critical problems for customer identity, authentication, and centralized systems. At the same time, PUF enables secured verification of delivery robots by customers for any robot.   
 
\subsection{Contributions}
This paper's significant contributions include: 
\begin{itemize}
    \item A \textit{novel framework} for robot based product delivery system.
    \item A \textit{first threat model} with DID access capabilities to the adversaries.
    \item A \textit{first} DID based privacy preserving customer verification scheme.
    \item A \textit{first} PUF based non deniable robot verification scheme for e-commerce industry.
    \item A rigorous Formal and Informal security analysis of the proposed work using \textit{tamrin prover} and \textit{security properties} respectively.
    \item A \textit{realtime} experimental outcomes and performance analysis.
\end{itemize}
In the proposed scheme, the robot requires much less storage (just one public key), less computation (just one signature verification and one encryption), and a very low communication overhead. To the best of our knowledge, this is the foremost work presented in this future-oriented research direction.  

\section{Background and Prelude}
In this section, we discuss basic preludes used and considered for designing of the proposed work. 

\subsection{Decentralized Identifier}

\begin{figure}[h]
  \centering
  \includegraphics[width=\linewidth]{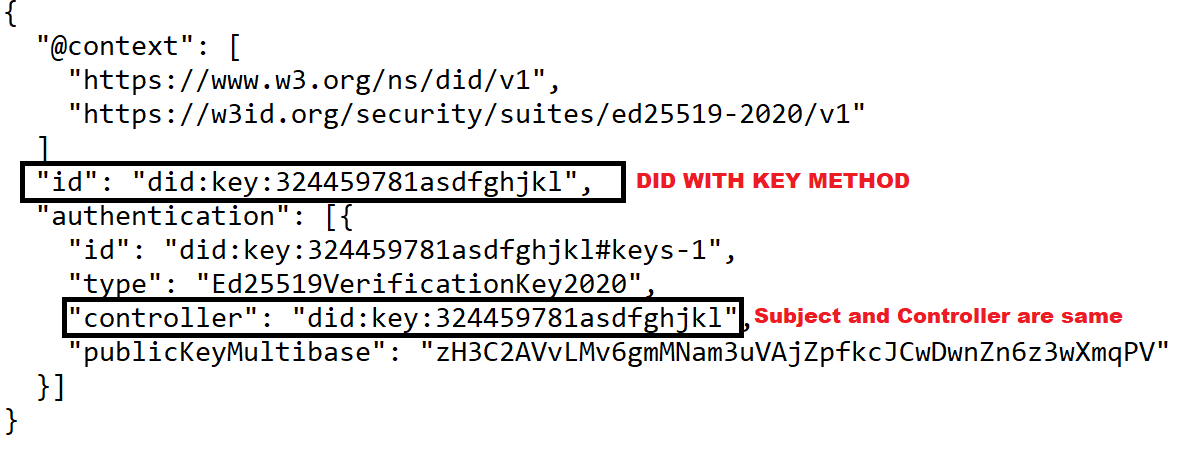}
 \caption{Resolved DID Document}
 \label{fig:DIDDoc}
\end{figure}

Decentralized Identifier (DID) \cite{DIAMIoT2020} provides a self-sovereignty to customers about their identities. Compared to a traditional approach in which customers can have unique identities, customers can have multiple identities and use those identities to escape from tracing by the intruder. We use the blockchain to store the DID documents in the proposed framework. These DID documents contain the mapping between DID and related metadata (e.g., order detail in the proposed framework). In the DID, there are holder, verifier and issuer entities. Trusted Issuer will issue verifiable credentials to holder and holder will present his/her DID and verifiable credential to the verifier. The verifier will retrieve the DID document from the ledger and verify the verification credentials. In the proposed scheme, we consider the government signed certificate as a verifiable credentials that proves name and birth date of the customer. Here government is trusted party and whose public key is available publicly for signature verification. Figure \ref{fig:DIDDoc} shows resolved DID document generated during customer verification phase. DID resolution involves three major entities: 1. Issuer (government here) who issues verifiable credentials and writes those credentials to the verifiable data registry. 2. Verifier (service provider here) who verifies the verifiable presentation sent by the customer. And 3. Holder (customer here) who receives verifiable credentials from different issuers and generates a verifiable presentation from those verifiable credentials based on the need of a verifier. Here customer decides which verifiable credentials are sufficient to show to the verifier. That's why DID is called Self-Sovereign Identity (SSI) also. 

In the proposed scheme, the customer stores its DID with order details in the DID document and registers these DID documents to the permissioned blockchain by executing the smart contract. A service enabler (e.g., the company's service provider server) in the proposed work needs to resolve DID requests by accessing DID documents. Access to DID documents can be only possible when the service provider server presents the customer's request (containing DID signed by their private key with valid metadata)  to the blockchain administrator. With the help of blockchain-enabled DID, we can achieve non-traceability, dynamicity, tamper-proof data storage, and full control of their own identity to the customer. Zero Knowledge Proof (ZKP) allows an entity to validate the truth of a statement without revealing any more information about the argument. With the help of ZKP, customer can present his/her VC to verifier with preserving the privacy of other details.   

\subsection{Blockchain}

A blockchain is a publicly distributed ledger that contains hash-connected blocks. Each block provides a tamper-proof environment to store the DID documents. With the help of permissioned blockchain, we can achieve decentralization, immutability, auditability, restricted access, traceable log generation and transparency. All nodes in a blockchain execute customer-defined functions called a smart contract to store the data. There are two types of blockchain that exists. One is permissioned (or private), and the other is a permissionless (or public) blockchain. Examples of permissioned blockchain include L3COS, Hyperledger fabric, Quorum, and R3, while Bitcoin and Ethereum are famous permissionless blockchain applications. In the proposed framework, we use a permissioned blockchain that the company's blockchain administrator maintains, and it has restricted access between the company's customer (who can store the DID) and the company's service provider server (who can resolve the DID into DID document) for verification purpose. 

\subsection{Physically Unclonable Function}
Physically Unclonable Function (a.c.a. PUF) is widely adopted hardware security approach for verification of devices. PUF works based on a consideration the each identical hardware chip produced by manufacturer provides a unique response to each challenge presented to it. PUF is proved unclonable, unpredictable and random function that takes challenge as input and provides device dependent response. Over the same PUF, if two device receives same challenge $C$ and those devices generate responses $R_1$ and $R_2$ respectively, then it is always true that $R_1$ $\neq$ $R_2$.

In the proposed scheme, company uses robot to deliver the product and each robot contains a hardware chip (we have used raspberry pi for experimental purpose) that provides unique challenge response pair. We consider that company who provides service to the customer stores thousands of these unique challenge-response pair for their robots and whenever customer orders the product, it randomly selects one challenge from the database (of robot who is going to provide service) and store it into customer mobile. 
Algorithm \ref{alg:3} presents steps of PUF based verification of robot (a.w.a customer also partially) in the intelligent robotic delivery system.
In this paper, we haven't consider that if robot is changed after order and before delivery. Solution to this we can consider as a future work.  
\begin{algorithm}
\caption{\textbf{: PUF Challenge Response Working}}
\label{alg:3}
\begin{algorithmic}[1]
\STATE Manufacturer generates Challenge Response Pair (CRP), 
\STATE Manufacturer securely store CRP database into cloud data base of Service Provider,
\STATE Service Provider Randomly selects challenge and store in to customer mobile device,
\STATE customer sends same challenge to the robot and receive the response,
\STATE customer generates verification request to service provider with signed responce
\STATE Service provider verifies challenge (send by customer) - Response pair (send by robot to customer). 
\STATE Service provider sends signed response to the customer with valid (0) or invalid (1).
\end{algorithmic}
\end{algorithm}

\subsection{Elliptic Curve Diffie Hellman Key Exchange and Signature Verification}
Elliptic Curve Cryptography (ECC) is a public key cryptography technique that provides equal security to Rivest-Shamir-Adleman (RSA) algorithm with very low computation complexity compare to it. In the proposed scheme, we use ECC to generate a public key and private key pairs of different entities involved in the proposed framework. With the help of Diffie hellman key exchange, entities share their keys securely \cite{patel2018internet}. In the proposed scheme, entities use the ECC generated public key for encryption purposes and the private key for signature purposes. An ECC based signature generation and verification enables us to implement the proposed scheme in any resource constraint devices. The algorithm \ref{alg:1} and algorithm \ref{alg:2} presents signature generation and signature verification steps using ECC.  
\begin{algorithm}
\caption{ \textbf{: ECC Signature Generation by Signer}}\label{alg:ECCSign}
\label{alg:1}
\begin{algorithmic}[1]
\STATE Compute $e_h$ = $SIP_h$(message), 
\STATE Generate $k_r$ where 1 $\leq$ k $\leq$ N,
\STATE Compute $G_p$($X_i$,$Y_i$)=$k_r$ * G, G is base point. 
\STATE Compute $r_i$ = $X_i$ (mod N). 
\STATE Compute $s_i$ = ($k_r^{-1}$)($h_i + r_i * d_A$) (mod N). $d_A$ is signer's secret key, $h_i$ is the fixed length left most bits of $e_h$.
\STATE Sign = ($r_i$,$s_i$)
\end{algorithmic}
\end{algorithm}
\begin{algorithm}
\caption{\textbf{: ECC Signature Verification by Verifier}}
\label{alg:2}
\begin{algorithmic}[1]
\STATE Compute $e_h$ = $SIP_h$(message), 
\STATE Compute $c_h$ = $s^{-1}$,
\STATE Compute $u_i$ = $h_i$ * $c_h$ (mod N),
\STATE Compute $u_j$ = $r_i$ * $c_h$ (mod N),
\STATE Compute $G_p$($X_i$,$Y_i$) = $u_i$ * G + $u_j$ * $Q_A$. Here $Q_A$ is public key of signer.
\STATE Return invalid if $G_p$($X_i$,$Y_i$) = 0
\STATE Return valid if $r_i$ = $X_i$ (mod N), else invalid
\end{algorithmic}
\end{algorithm}
\subsection{SIPHash : A Key based Hash Function}
A hash function conducts a one-way calculation on a piece of data, resulting in a series of bytes that, one believes, is entirely unpredictable and cannot be employed to deduce the original data. SIPHash was proposed by  Jean-Philippe Aumasson and Daniel J. Bernstein that is key-based hash function and significantly faster than other hash functions. It is also proven to be secured against hash-collision attacks and other attacks over hash functions. For agreed 128 bit key $K_i$ between customer $U_i$ and $U_j$ and message $M_1$, $U_i$ generates 64 bit message authentication code (MAC) SIPHash as $SIP_h$($M_1$, $K_i$). SIPHash assures that, having $M_1$ and $SIP_h$($M_1$, $K_i$), an attacker who is unaware of $K_i$ cannot compute $K_i$ or $SIP_h$($M_2$, $K_i$) any message $M_2$ $\notin$ {$M_1$}. We have used SIPHash for computation time optimization in hash computations and to achieve collision resistance as well as unforgeability. Readers can also use one way hash functions such as SHA-256 in place of SipHash based on their application. 
\section{Contributed Work}
In this section, we present a detailed discussion of our contribution. First, we discuss a novel proposed framework in which we present a novel DID-based framework to verify a customer-based self-sovereign identity generated/controlled customers themselves. And later, We discuss threat model to the contributed work. Next, we discuss the customer verification approach in which the robot will get assurance that the receiving customer is a customer of the company. We discuss the robot verification approach in which customers will get assurance that the robot delivering a product is the same robot sent by a company. 
\subsection{Proposed Framework}
\begin{figure}[!hbt]
  \centering
  \includegraphics[width=\linewidth]{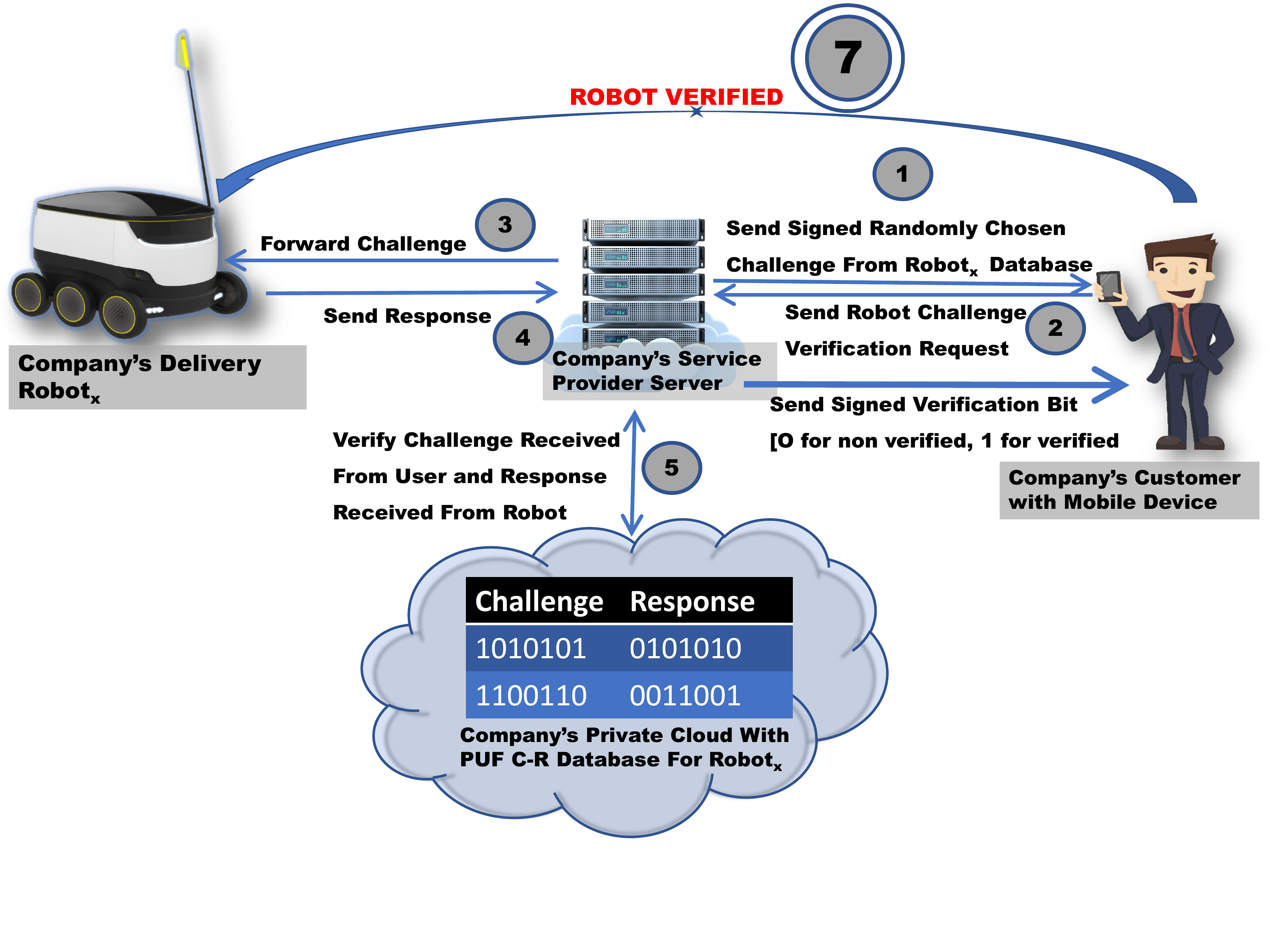}
 \caption{Proposed DID Framework}
 \label{fig:framework}
\end{figure}

Decentralized Identifier (DID) is a future of identity derivation for the billions of customers and devices. DID involves numerous methods based on customer requirements and the central registry used. DID empowers customers to create and control their own identity, which is neither centralized nor revocable. As shown in Fig. \ref{fig:framework}, we propose a DID-based framework for customer verification for the e-commerce company. To the best of our knowledge and analysis, this framework is highly applicable in many other services where the customer is empowered for its own identity, and the service providers want to verify the customer based on that identity. \textbf{Verifiable Credentials (VC)} are used for creating a trust-building among holder/customer and verifier/service provider. customers can receive multiple VCs from numerous issuers (i.e., government, university, etc.), and based on the requirement of the service provider, and use generates \textbf{Verifiable Presentation (VP)} from VCs and forwards VP to the service provider for verification. The service provider verifies the VP based on the issuer's signature on each VC.   

The proposed framework consists of seven basic steps from DID creation to DID and VC verification by the e-commerce service provider. We consider permissioned blockchain managed by the company's Blockchain Administrator (BA) in the proposed framework. This blockchain is accessible only by the registered customer and the company's authorized entities, such as the service provider. The most important advantage of using permissioned blockchain is that the BA only allows verified issuers to register their DID and completely secures system from \textbf{fake issuers} and \textbf{fake VC} attacks in DID. The proposed framework consists of following steps :
\begin{enumerate}
    \item Customer creates a DID \textbf{(step 1)} with hashed value of order details and stores this DID in to permissioned blockchain of the company \textbf{(step 2)}.
    \item Now, customer presents signed DID (using own private key) and signed VCs (signed using issuer's (i.e. government agency's) private key) to the company's service provider server \textbf{(step 3)}.
    \item Company's service provider starts DID verification phase and provides a DID to the permissioned blockchain \textbf{(step 4)}, resolves a DID into DID Document \textbf{(step 5)} and receives this resolved DID Document \textbf{(step 6)}. Here, company's service provider will receive two DID documents. First DID document consists of public key of issuer and second DID document consists of public key of customer and hashed value of order details. Here, we considered that signed VCs of the customer also contains DID information about issuer that helps server to resolve the DID of issuer and permissioned blockchain of the company consists a DID information about all valid government issuers.
    \item Company's service provider verifies DID and order details of the customer using public key received from the DID document of customer and validates VCs using public key derived from the DID document of the issuer \textbf{(step 7)}. 
\end{enumerate}

\subsection{Threat Model}
In this subsection, we consider two types of the threat model. As the first type of threat model, we consider an adversary with DID-related access and another adversary with access to all public channels and can receive, modify, and delete the communicated messages. To the best of our knowledge, this is the first threat model that provides some power to the adversaries related to DID accesses. 

\noindent \textbf{Type 1 Adversary $\xrightarrow{}$} Adversary with DID and VCs access capabilities: 
\begin{itemize}
    \item The adversary can access the permissioned blockchain.
    \item Customers' privacy can leak if an adversary can access order data in plain text, so the anonymity of customer order details is a key challenge in managing customer privacy. 
    \item The adversary has access to the signed VCs of the customer.
    \item An adversary can trace all the communication if the same DID is used for multiple orders. 
\end{itemize}

\noindent \textbf{Type 2 Adversary $\xrightarrow{}$} Adversary with communication channel access capabilities: 
\begin{itemize}
    \item The adversary can eavesdrop, modify, and delete all communication over a public channel between the customer, service provider, and robots. 
    \item Here we consider that service providers can have an insider, the customer may try to fake order data, and robots may try to cheat the company by selling products to another customer.
\end{itemize}
\subsection{Proposed Scheme} 
In this subsection, we propose a novel scheme for customer verification and robot verification for the e-commerce company.  The proposed scheme consists of three major phases. \textbf{1:} Customer Registration Phase \textbf{2:} Customer Verification Phase and \textbf{3:} Robot Verification Phase. The customer verification phase consists of two sub phases : \textbf{2.1:} Customer DID Verification Phase and \textbf{2.2:} Customer VCs Verification Phase.
\begin{figure}[!hbt]
  \centering
  \includegraphics[width=\linewidth]{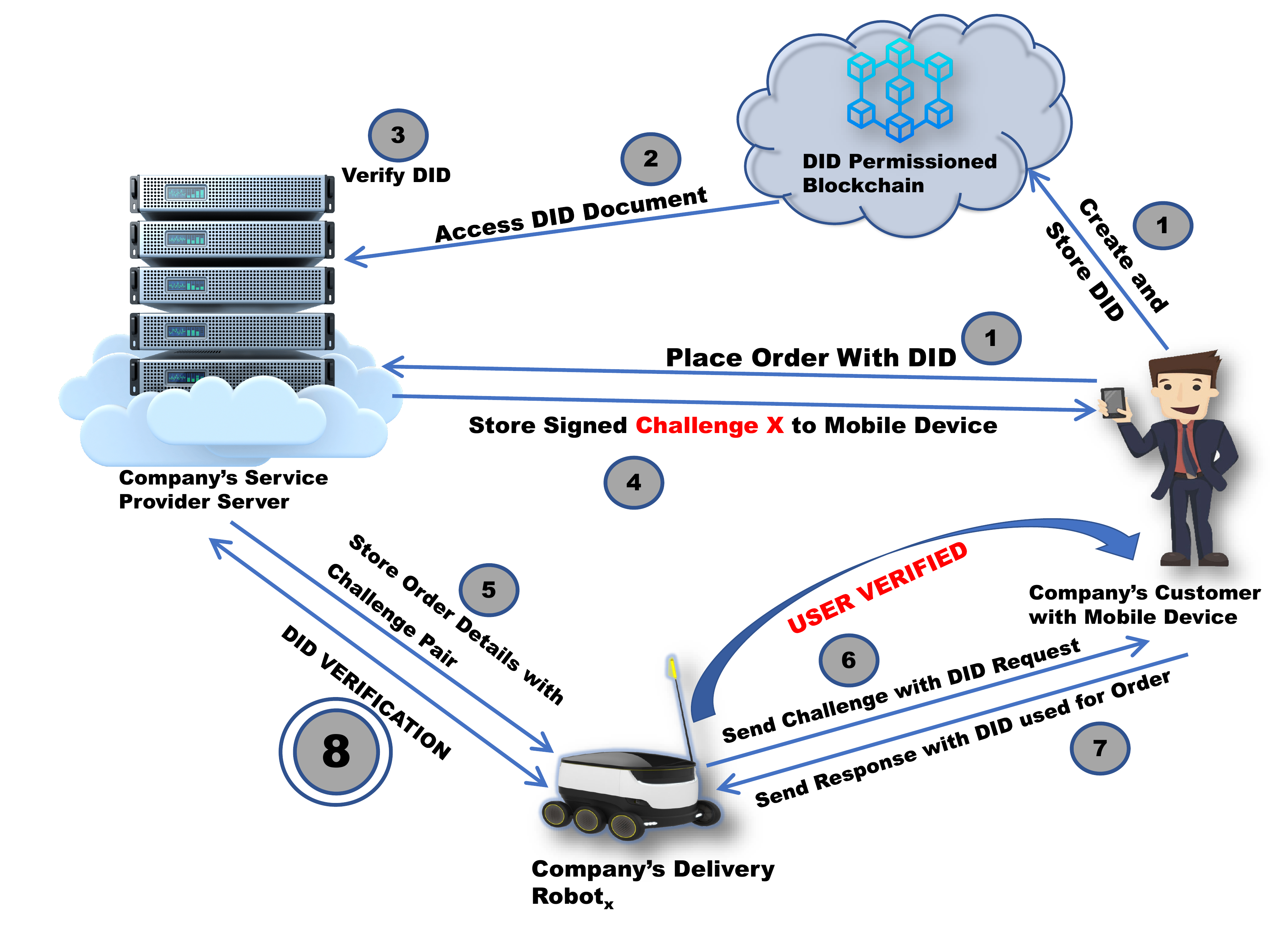}
 \caption{Proposed Customer DID Verification}
\end{figure}
\subsubsection{\textbf{Customer Registration Phase}} The customer registration phase consists of two steps. In the first step, the customer obtains verifiable credentials (VCs) from the government issuers, and in the second step, a customer obtains VCs from the service provider company (valid only for a day) by presenting the VCs obtained from the government issuers. As per the policy of an e-commerce company, customers can merge multiple VCs and create a verifiable presentation (VP) to prove the eligibility for item purchase (i.e., country (to assure location), date of birth (to assure customer not buying the acidic item if they are below some age)). For the \textbf{Step 1}, we assume that the customer has already obtained valid VCs from the government issuers and they have stored them as a $(VC)^{pr}_{Gvt}$, the service provider has $DID_{Gvt}$ and blockchain have entries for the public key for government issuers \cite{GopeDID2022}. Here, each company's service provider is also having a $(VC)^{pr}_{R_{comp}}$ that is associated with $DID_{comp}$ and verification of this verifiable credential will prove that it is valid service provider of the company.  

\noindent \textbf{Step 2: Obtaining VCs from the Company's Service Provider (CSP) : }  

\textbf{1.} The customer generates random number $n_u$ and send message $M_1$ = \{$n_u$, $(VC)^{pr}_{Gvt}$\} to the company's service provider. In message $M_1$, $n_u$ plays a role of challenge by the customer to CSP.  

\textbf{2.} Upon receiving $M_1$, CSP verifies VCs issued by government issuers. After successful verification, The CSP generates $DID_{csp}$ (separate for each new customer registration), public key $V_{csp}$ and private key $R_{csp}$. The CSP binds $DID_{csp}$ and public key $V_{csp}$ in permissioned blockchain. Now CSP generates random nonce $n_{csp}$ and generates VC for customer as $VC_u$ = $SIP_h$($n_{csp}||n_u||(VC)^{pr}_{Gvt}$). Furthermore, CSP signs this credential as $(VC_u)^{pr}_{R_{csp}}$ and response to challenge as $(n_u)^{pr}_{R_{csp}}$ using its private key. The CSP also receives customer specific temporary $TID_u$ and credential $cr_u$ from the blockchain administrator. These two parameters are only used during registration step and valid only for one time use. Here, the e-commerce company can decide whether to provide One Time Credential (OTC) or One time Link (OTL) for registering DID in the blockchain.  The CSP creates a message $M_2$ = \{$(n_u)^{pr}_{R_{csp}}$, $(VC_u)^{pr}_{R_{csp}}$, $DID_{csp}$, $n_{csp}$, $TID_u$, $cr_u$\} and sends to customer.       

\textbf{3.} Upon receiving message $M_2$ from CSP, customer verifies response to challenge by resolving DID of CSP. The customer generates $DID_{u}$ (separate for each new ), public key $V_{u}$ and private key $R_{u}$. Customer binds $DID_{u}$ and public key $V_{u}$ with permissioned blockchain using pair \{$TID_u$, $cr_u$\}. Now customer creates a response $(n_{csp})^{pr}_{R_{u}}$ and sends message $M_3$ = \{$(n_{csp})^{pr}_{R_{u}}$, $DID_{u}$\} to the CSP. Customer stores \{$(VC_u)^{pr}_{R_{csp}}$, $DID_{u}$, $DID_{csp}$, $R_{u}$\} in his/her secret memory of wallet.   

\textbf{4.} Upon receiving message $M_3$ from customer, CSP resolves $DID_{u}$ and verifies response to challenge $(n_{csp})^{pr}_{R_{u}}$ using $V_{u}$. After successful verification, CSP gets assurance that customer has registered his/her DID as well as has obtained VCs. 

\begin{table*}
    \centering
    \caption{Customer Verifiable Credential Obtaining Phase}
    \begin{tabular}{|lcl|} \hline
    \begin{minipage}{.3\textwidth}
      \includegraphics[width=10mm, height=10mm]{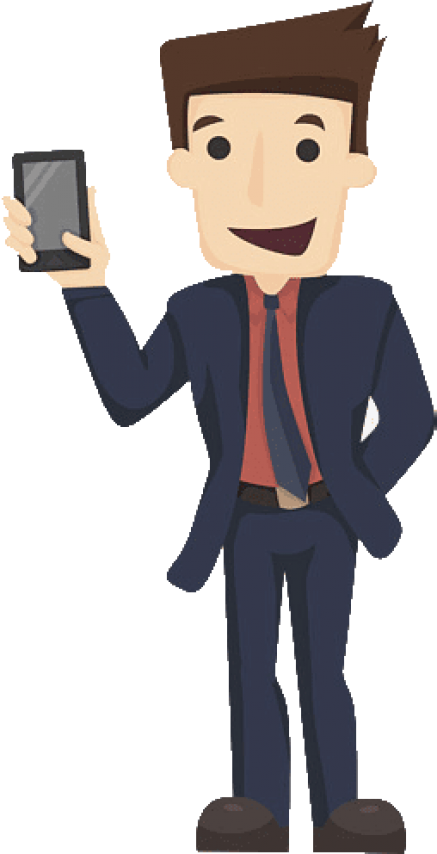}
    \end{minipage} & & \begin{minipage}{.3\textwidth}
      \centering \includegraphics[width=10mm, height=15mm]{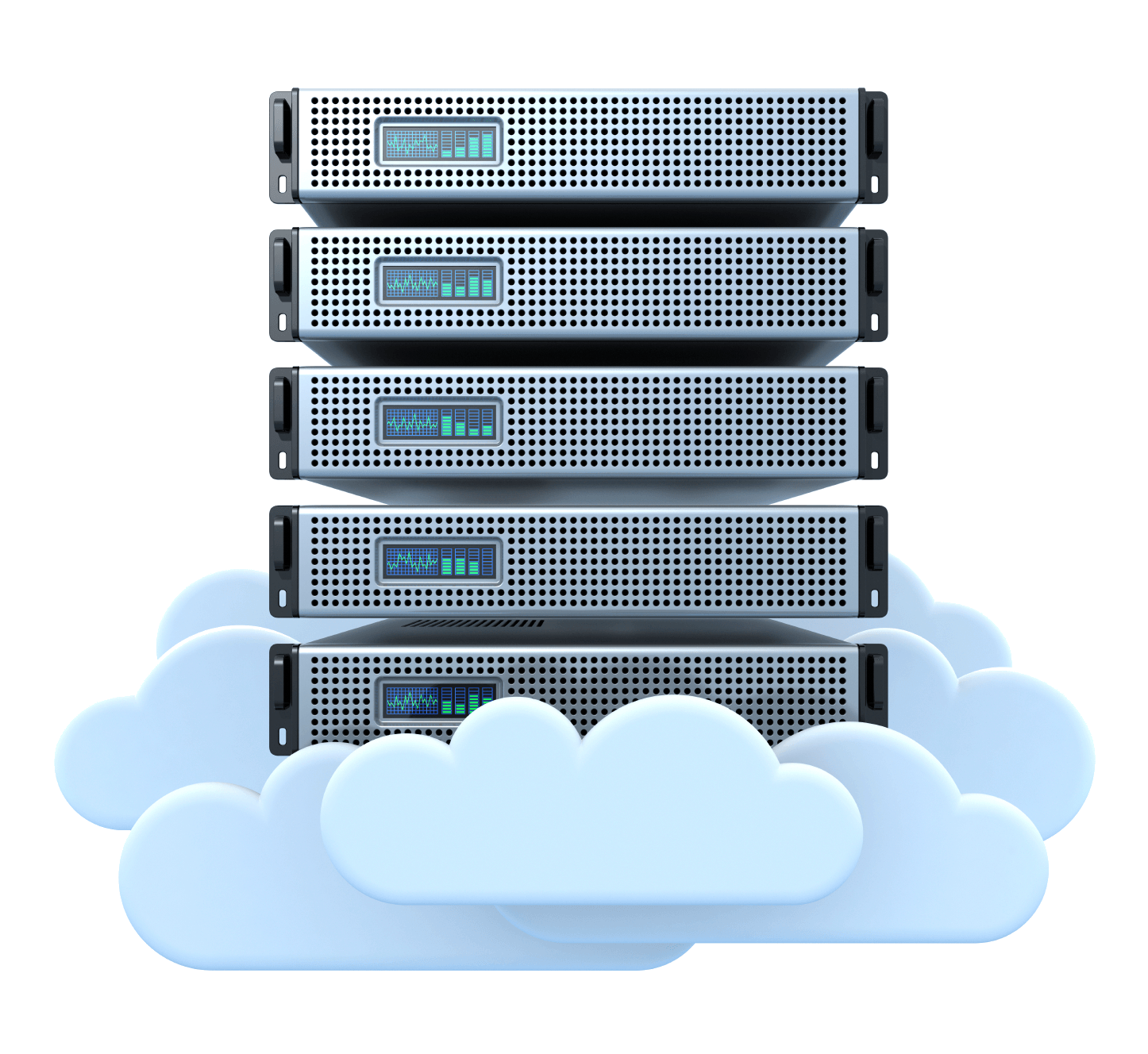}
    \end{minipage} \\
    \hspace{0.1cm}\textbf{Customer} & & \hspace{1.5cm}\textbf{Company Service Provider}  \\ \hline
    Generate $n_u$,  &  &   \\
    $M_1$ = \{$(VC_u)^{pr}_{Gvt}$, $n_u$\}, & & \\
   & \framebox[1.1\width]{$\xrightarrow{M_1}$} &  \\ 
   &  & Verify $(VC_u)^{pr}_{Gvt}$, \\
   &  & Generate $DID_{csp}$,  $V_{csp}$, $R_{csp}$, \\
   &  & Bind \{$DID_{csp}$,  $V_{csp}$\} to blockchain, store $R_{usp}$ in secret memory, \\
   &  & Generate $n_{csp}$, \\
   &  & Generate $VC_{u}$ = $SIP_h$($n_u|| n_{csp}||(VC_u)^{pr}_{Gvt}$),\\
   &  & Sign $(VC_{u})^{pr}_{R_{at}}$, $(n_u)^{pr}_{R_{usp}}$, \\
   &  & $X_1$ = $SIP_h$($salt||(VC_{U})^{pr}_{R_{at}}$),\\
   &  & Generate  $TID_u$, $cr_u$,\\
  Generate $DID_{u}$,  $V_{u}$, $R_{u}$,  &  & $M_2$ = \{$(n_u)^{pr}_{R_{usp}}$, $(VC_u)^{pr}_{R_{usp}}$, $DID_{csp}$, $n_{csp}$, $TID_u$, $cr_u$\},\\
  Bind \{$DID_{u}$,  $V_{u}$\} to blockchain using \{$TID_u$, $cr_u$\}, & \framebox[1.1\width]{$\xleftarrow{M_2}$} & \\
  store $R_{u}$ in secret memory, & &   \\
  Sign $(n_{csp})^{pr}_{R_{u}}$, & & \\
  $M_3$ = \{$(n_{csp})^{pr}_{R_{u}}$, $DID_{u}$\}, & &  \\  
 Keep \{$DID_{u}$, $DID_{csp}$\} & \framebox[1.1\width]{$\xrightarrow{M_3}$} & Resolves $DID_{u}$ and verifies $(n_{csp})^{pr}_{R_{u}}$, \\
 Store \framebox[1.1\width]{$(VC_u)^{pr}_{R_{csp}}$},  & & \\
  &  &  \\ \hline
    \end{tabular}
    \label{tab:1}
\end{table*}

\subsubsection{\textbf{Customer and Order Verification Phase}} The customer verification phase consists of two-steps. In the first step, customer places an order with e-commerce company's service provider and receives order specific verifiable credentials (OSVCs), and in the second step, the company's robot verifies the customer, order details and OSVCs and deliver a product to the verified customer.   

\noindent \textbf{Step 1 $\xrightarrow{}$ Order Placing Phase by customer:} 

\textbf{1.} First, the customer picks the Product ID as $PID_i$ (that is part of the global trade item number and is unique for each product) and product details (including ongoing price) as $PD_i$ for that product ID from the company's database. It computes $X_1$ = $SIP_h$($PD_i || PID_i$). Next Customer generates random nonce $n_u^*$  and sign this nonce using ECDSA as $S_1$ = $sign(n_u^*)^{Pr}_{R_u}$. Now customer creates a message $M_1$ = $Enc(X_1,S_1,PID_i,(VC_u)^{pr}_{R_{csp}},DID_u)^{Pub}_{V_{csp}}$ and sends \{$M_1$, $T_S$\} to the CSP where $T_S$ is the current timestamp. Here, the customer obtains a public key of the CSP by resolving DID of the CSP that is available with him. We have used timestamp verification for two purposes. One is to prevent replay attacks, and the other is to ensure that customers can order products from any time zone.

\textbf{2.} After receiving message \{$M_1$, $T_S$\} from the customer, CSP verifies timestamp and time synchronization. Furthermore, CSP decrypts $M_1$ using its private key and retrieves $\{X_1||S_1||PID_i|| (VC_u)^{pr}_{R_{csp}} \\ || DID_u\}$. First, it resolves customer DID and retrieves $n^*_u$ after verifying signature $S_1$. Next, CSP retrieves product details as $PD_i^*$ using $PID_i$ and verifies $X_1^*$ = $SIP_h$($PD_i^* || PID_i$) $\stackrel{?}{=}$ $X_1$. Now, CSP verifies $(VC_u)^{pr}_{R_{csp}}$ using own public key and and validates eligibility of customer based on presented VCs and product details. After these all validation, CSP generates $n^*_{csp}$, generates OSVC = $SIP_h$($n^*_{csp} || n^*_u || X_1^*$) and signs it as $S_2$ = $Sign(OSVC)^{pr}_{R_{csp}}$. Furthermore CSP generates $S_3$ = $Sign(n^*_{csp})^{pr}_{R_{csp}}$, $M_2$ = $Enc(S_2,S_3)^{pub}_{V_{u}}$ and send \{$M_2$, $T_S$\} to customer.    

\textbf{3.} Upon receiving \{$M_2$, $T_S$\}, the customer verifies timestamp, decrypts $M_2$, and retrieves \{$S_2$,$S_3$\}. Customer verifies $S_3$ using a CSP public key that customer has retrieved after resolving DID of CSP. Now, the customer computes $X_2$ = $SIP_h$($n^*_{csp}||n^*_{u}$) and generates order specific verifiable presentation \textbf{$OSVP_1$ = \{$S_2$, $X_2$\}}. Customer will use this presentation while receiving a product.     

After this step, CSP selects $Robot_x$ for the delivery and stores order specific public key of CSP ($V_{csp}$) in secret memory (that is secure element) of the $Robot_x$. The CSP also maintains \{$DID_u$, $Robot_x$\} pair.

\noindent \textbf{Step 2 $\xrightarrow{}$ Customer, order details and OSVPs verification Phase by company's robot:} This step starts when robot reaches for product delivery and 

\textbf{1.} Robot generates a request towards the customer to present the valid challenge, OSVC for to receive the product. 

\textbf{2.} Customer presents $M_3$ = $Enc(OSVP_1, DID_u, sign(n^*_{u})^{pr}_{R_{u}})^{pub}_{V_{csp}}$ to the robot. Robot forwards received message $M_3$ to the CSP. 

\textbf{3.} CSP decrypts $M_3$ and retrieves  \{$S_2$, $X_2$, $DID_u$, $sign(n^*_{u})^{pr}_{R_{u}}$\}. CSP resolves customer DID and verifies signature $sign(n^*_{u})^{pr}_{R_{u}}$. Next CSP performs $X_2^*$ = $SIP_h$($n^*_{csp}||n^*_{u}$) $\stackrel{?}{=}$ $X_2$. After this two verification, CSP verifies $S_2$ that is signed order specific verifiable credential and contains order details. If the $S_2$ verification succeed, it generates success signal $Succ_{sign}$ and if it is failed, it generates failure signal $Fail_{sign}$. Later it signs it as a $S_5$ = $Sign(Succ_{sign})^{pr}_{R_{csp}}$ or $S_5$ = $Sign(Fail_{sign})^{pr}_{R_{csp}}$ and sends back to the robot. Based on stored public key of the CSP, robot verifies $S_5$ and decide to deliver a product or not. 

\begin{table*}[hbt!]
    \centering
     \caption{Order Placing and Customer Verification Phase}
    \begin{tabular}{|lll|} \hline
    \begin{minipage}{.3\textwidth}
      \centering \includegraphics[width=10mm, height=15mm]{Customer.png}
    \end{minipage}&\begin{minipage}{.3\textwidth}
      \centering \includegraphics[width=10mm, height=15mm]{CompanyServiceProvider.png}
    \end{minipage}  & \begin{minipage}{.3\textwidth}
      \centering \includegraphics[width=10mm, height=15mm]{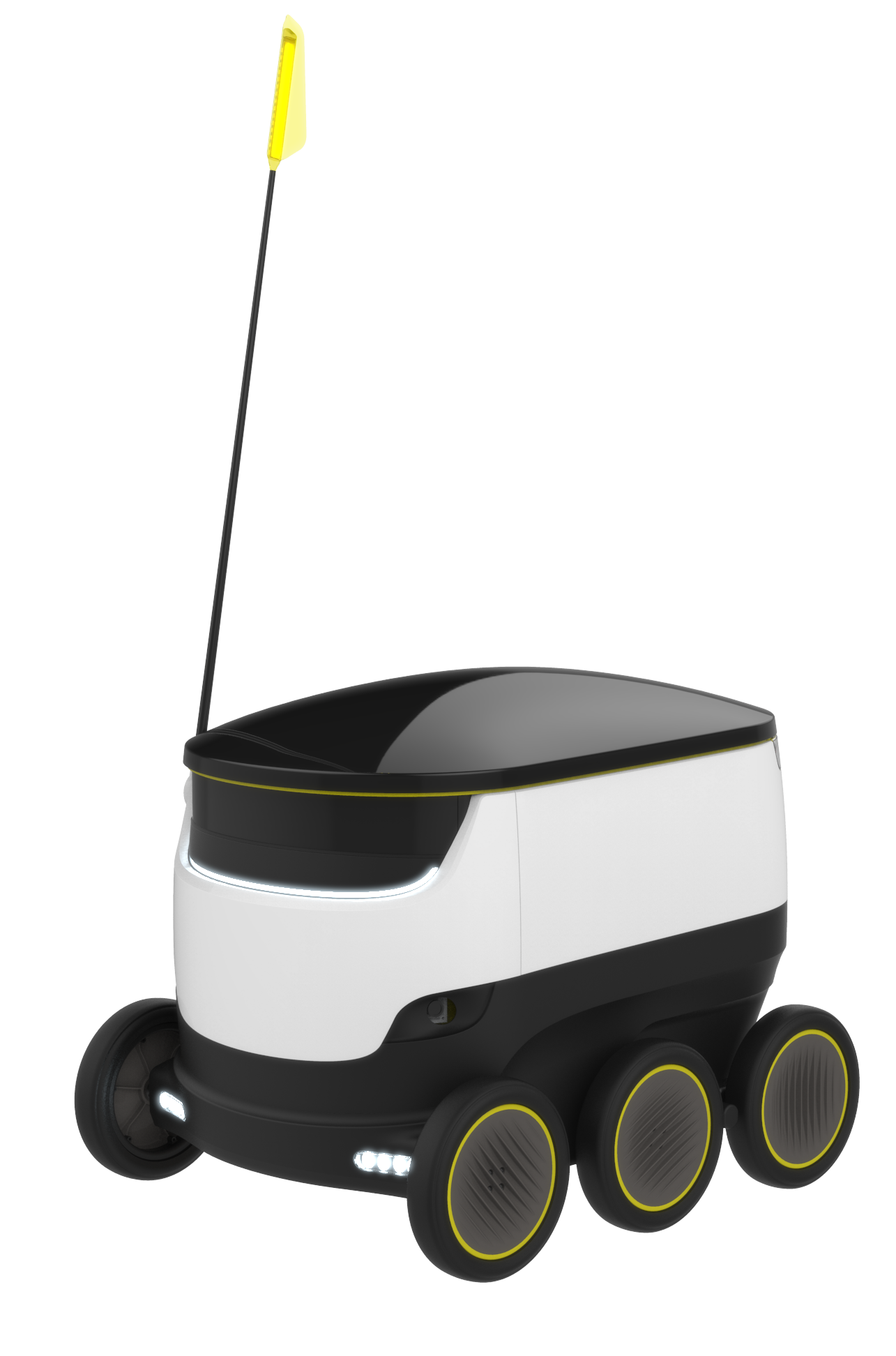}
    \end{minipage}  \\ 
    \hspace{2cm}\textbf{Customer} & \hspace{1.2cm}\textbf{Company Service Provider} & \hspace{2.2cm}\textbf{Robot}  \\ \hline
    \textbf{Step 1 $\xrightarrow{}$ Order Placing:} & & \\ \hline
    Pick $PID_i$, $PD_i$, &  & \\
    $X_1$ = $SIP_h$($PD_i || PID_i$), & & \\
    Generate $n_u^*$, $S_1$ = $sign(n_u^*)^{Pr}_{R_u}$, &  & \\
    $M_1$ = $Enc(X_1,S_1,PID_i,(VC_u)^{pr}_{R_{csp}},DID_u)^{Pub}_{V_{csp}}$, & & \\
    \hspace*{0pt}\hfill \framebox[1.1\width]{$\xrightarrow{M_1, T_S}$}, & Verify $T_S$ & \\
    & Resolves $DID_u$, verify $S_1$, & \\
    & Get $PD_i^*$ using $PID_i$, & \\
    & Verify $X_1^*$ = $SIP_h$($PD_i^* || PID_i$) $\stackrel{?}{=}$ $X_1$, & \\
    & Verify $(VC_u)^{pr}_{R_{csp}}$, & \\
    & Generates $n^*_{csp}$,  & \\
    & Generates OSVC = $SIP_h$($n^*_{csp} || n^*_u || X_1^*$), & \\
  Verify $T_S$,  & $S_2$ = $Sign(OSVC)^{pr}_{R_{csp}}$, & \\
  Decrypt $M_2$,  &  $S_3$ = $Sign(n^*_{csp})^{pr}_{R_{csp}}$, & \\
  Verify $S_3$,  & $M_2$ = $Enc(S_2,S_3)^{pub}_{V_{u}}$, & \\
  Compute $X_2$ = $SIP_h$($n^*_{csp}||n^*_{u}$),   & \framebox[1.1\width]{$\xleftarrow{M_2, T_S}$} & \\
  \textbf{$OSVP_1$ = \{$S_2$, $X_2$\}}, & \framebox[1.1\width]{$\xrightarrow{V_{csp}}$} & \\   
  & Keep \{$DID_u$, $Robot_x$\}, & \\ \hline
  & & \hspace*{0pt}\hfill \textbf{Step 2 $\xrightarrow{}$ Customer verification by robot:}\\ \hline
  $\xrightarrow[{M_3 = Enc(OSVP_1, DID_u ,sign(n^*_{u})^{pr}_{R_{u}})^{pub}_{V_{csp}}}]{To Robot_x}$ &  & $\xleftarrow{Request to customer}$ \\
  & & $\xleftarrow{{M_3 to CSP}}$\\
  & Decrypt $M_3$, & \\
  & verifies $sign(n^*_{u}){pr}_{R_{u}}$, & \\
  & $X_2^*$ = $SIP_h$($n^*_{csp}||n^*_{u}$) $\stackrel{?}{=}$ $X_2$, & \\
  & Verify $S_2$, & \\
  & Generates $Succ_{sign}$, & \\
  & $S_5$ = $Sign(Succ_{sign})^{pr}_{R_{csp}}$, & \\
  & \framebox[1.1\width]{$\xrightarrow{S_5}$} & Verify $S_5$ and signal, \\
  &  & \\ \hline
    \end{tabular}
    \label{tab:my_label}
\end{table*}
\subsubsection{Robot Verification Phase} 
\begin{figure}[!hbt]
  \centering
  \includegraphics[width=\linewidth]{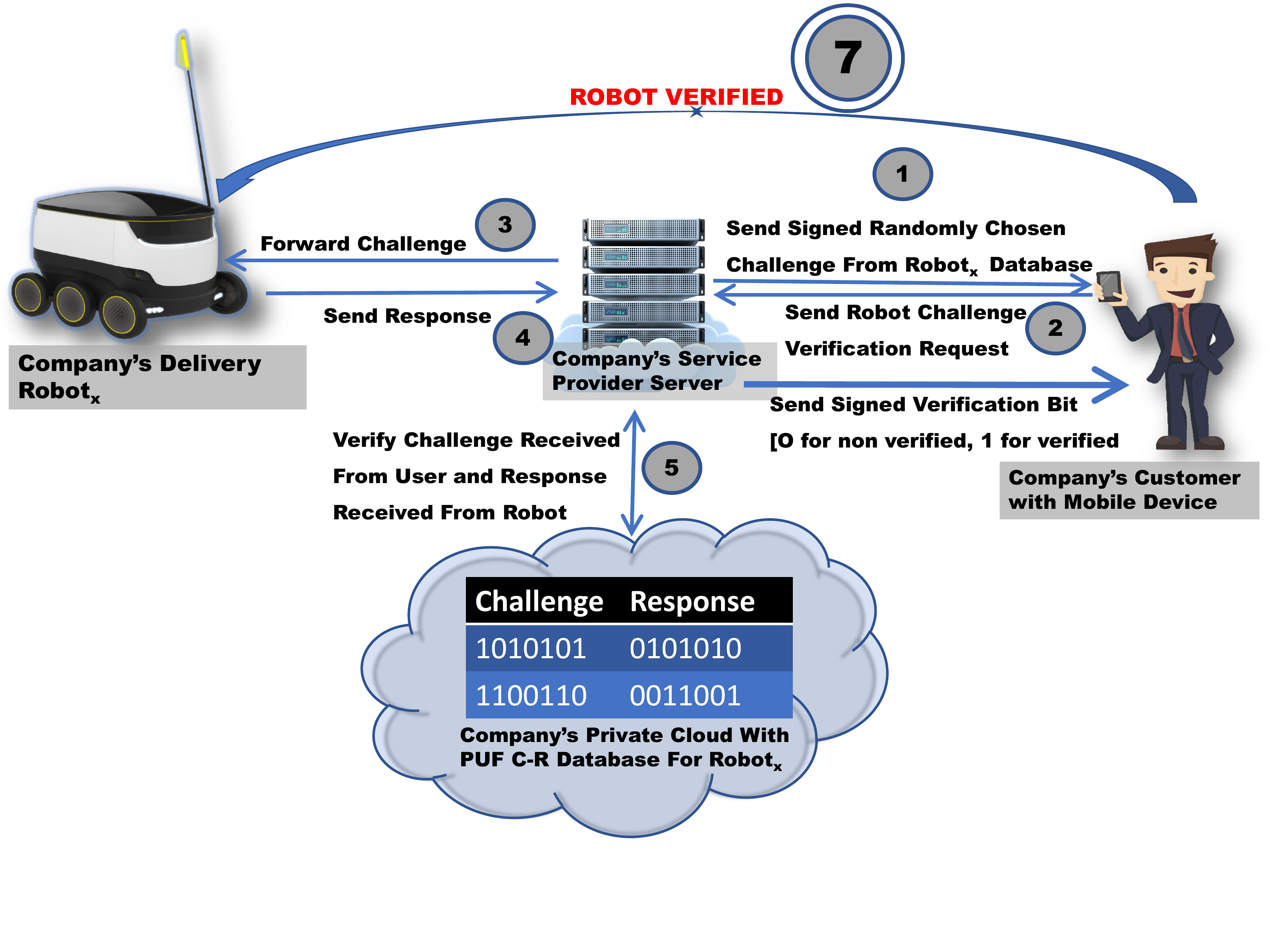}
 \caption{Proposed Robot Verification Approach}
 \label{Fig:Robot_Veirifcation}
\end{figure}

The robot verification phase can transpire either in parallel to customer verification, or it can also transpire before or after the customer verification phase. In the proposed work, we have considered this step after customer verification step.  Here, we have considered that CSP maintains a PUF Challenge-Response Pair (CRP) record for each in-service robot. As per shown in Fig. \ref{Fig:Robot_Veirifcation}, company's private cloud maintains CRP for $Robot_x$. This database is accessible only by the admin team of CSP. The robot verification phase consists of two steps. Step one is associated with challenge acquisition in which the customer will request to CSP for the challenge, and CSP will select and provide any random challenge from the CRP database of $Robot_x$. In step two, the customer will perform challenge verification through CSP. Here, we have not considered direct robot verification by the customer to control the customer's access to the robot. The E-commerce company can decide to handle this verification through CSP or directly between customer and robot.  

\noindent \textbf{Step 1 $\xrightarrow{}$ Challenge Acquisition by Customer:} 

\textbf{1.} The Customer Mobile Agent (CMA) generates the $Robot_{verify}$ signal, $n^{""}$ and sends it to CSP. Here we have considered that the customer is unaware of any information about the robot. The CMA generates message $M_1$  = $Enc((VC_u)^{pr}_{R_{csp}}, Sign(n^{""})^{pr}_{R_{u}}, DID_{u}, Robot_{verify})^{pub}_{V_{csp}}$ and sends \{$M_1$, $T_S$\} to CSP.

\textbf{2.} Upon receiving \{$M_1$, $T_S$\}, CSP performs time stamp validation and later decrypts $M_1$ using its own private key. Next, CSP resolves $DID_{u}$ and verifies signature of $n^{""}$. Furthermore, CSP also validates $VC_u$ using its public key. After these validations, CSP randomly chooses $Challenge_i$ from the CRP database of the $Robot_x$. The CSP generates a message $M_2$  = $Enc(Sign(Challenge_i)^{pr}_{R_{csp}})^{pub}_{V_{u}}$ and sends \{$M_2$, $T_S$\} it to the CMA.     

\textbf{3.} Upon receiving \{$M_2$, $T_S$\}, CMA validates time stamp and decrypts $M_2$ using his/her own private key. The CMA validates signature of $Challenge_i$ upon resolving $DID_{csp}$ and uses $Challenge_i$ for next step.  

\noindent \textbf{Step 2 $\xrightarrow{}$ Challenge Verification by Customer:} 

\textbf{1.} The CMA signs $Challenge_i$ as $S_1$ = $Sign(Challenge_i)^{Pr}_{V_{u}}$ and sends \{$S_1$, $T_S$\} to CSP. 

\textbf{2.} Upon receiving \{$S_1$, $T_S$\}, the CSP validates timestamp and signature $S_1$. The CSP resigns  $Challenge_i$ as $S_2$ = $Sign(Challenge_i)^{Pr}_{R_{csp}}$ and sends it to the $Robot_x$. The $Robot_x$ verifies signature $S_2$ using ${Pub}_{V_{csp}}$ and generates $M_3$ = $Enc({Challenge_i, Response_i})^{Pub}_{V_{csp}}$. The $Robot_x$ sends \{$M_3$, $T_S$\} to the CSP.    

\textbf{3.} After validating timestamp, the CSP decrypts $M_3$ and retrieves \{$Challenge_i$, $Response_i$\} pair. After that CSP validates this pair with CRP database. Upon successful/failed validation, the CSP generates verification bit $VB_{1}$ (for successful) or $VB_{0}$ for failure and sign it as  $S_3$ = $Sign(VB_{1})^{Pr}_{R_{csp}}$ and sends \{$S_3$, $T_S$\} to the CMA. 

\textbf{4.} The CMA validates the timestamp and signature of the verification bit using public key of CSP received upon resolving $DID_{csp}$. Based on the verification bit, CMA informs the customer to receive or decline the delivery. 

\section{Security Verification}
\subsection{Formal Security Verification Using Tamarin Prover}
In this section, we present formal security proof for the proposed approaches using the tamarin prover. First we present security proof for the customer verification scheme followed by robot verification approach. To the best of our knowledge, this is the first formal security verification for the customer verification system that uses decentralized identifier. 

A security protocol verification tool called the Tamarin prover allows both data manipulation and unrestricted verification in the metaphorical model \cite{meier2013tamarin}. Tamrin prover is a robust asynchronous tool for security protocol design and evaluation. It takes as input a proposed security protocol, operations performed by agents running the proposed protocol (e.g., the customer, the robot, and the trusted company's service provider), a delineation of the attacker, and a delineation of the properties of the proposed protocol. In tamarin prover, by manipulating network messages and making new ones, the attacker and the protocol communicate with each other. For to prove the security of proposed protocol, we have used some built in cryptographic primitives of the tamarin prover and representation of those functions are as follows in our analysis. 1. For denoting asymmetric encryption : \textbf{\textit{aenc}}, 2. For denoting asymmetric decryption : \textbf{\textit{adec}}, 3. For cryptographic hash function : \textbf{\textit{h}}, 4. For public key : \textbf{\textit{pk}}, 5. For private key : \textbf{\textit{k}}, 6. For signature generation : \textbf{\textit{signg}}, 7. For signature generation : \textbf{\textit{signv}}. We can model the properties of these functions using \textit{equations}. An attacker in the tamarin prover is by default a Dolev-Yao attacker who can intercept the network traffic and update, remove, and inject a message over a network. However, it can not break the cryptographic primitives, such as breaking the encryption (ensuring confidentiality) and changing the signature (ensuring integrity).  

A state of the system is represented by a multiset of facts that store the state's information. The multiset rewriting rules used for modeling the protocol operates states. Ex. \textbf{\textit{$out(m)$}} and \textbf{\textit{$in(m)$}} represents that message m is forwarded over public channel and message m is received from the public channel respectively. Each rule is consists of set of premises (\textit{P}), set of operations (\textit{O}) and set of conclusions (\textit{C}) those we can represent as : \[ \textit{P} -  [O] \xdashrightarrow{} C \] 

The conclusion's facts will be added to the state while the premises are removed to carry out the rule, which demands that all of the premise's facts be present in the current state. In tamarin, $\sim$V, $\$V$, $\#V$, !F denotes that V is a fresh variable, V is a public variable, and V is a temporal time-sensitive variable, and F is a persistent fact respectively.

Now, with the help of \textit{rules}, we will model the proposed protocol. In the proposed protocol, we have defined following rules:
\begin{itemize}
    \item \textit{PLATFORM\_SETUP}
    \item \textit{ISSUER\_SETUP}
    \item \textit{VERIFIER\_SETUP}  
    \item \textit{VC\_OBTAIN}
    \item \textit{ORDER\_PLACING}
    \item \textit{CUSTOMER\_VERIFY}
\end{itemize}
With the help of \textit{lemmas}, we will prove the important security properties achieved in it. We have considered following lemmas:
\begin{itemize}
    \item \textit{Anonymity}
    \item \textit{Privacy}
    \item \textit{Authentication}
    \item \textit{Verification}
    \item \textit{can\_be\_revoked}
    \item \textit{signature\_nonlinkability}
\end{itemize}
\begin{figure*}
    \centering
    \includegraphics[width=\linewidth,height=4.75cm]{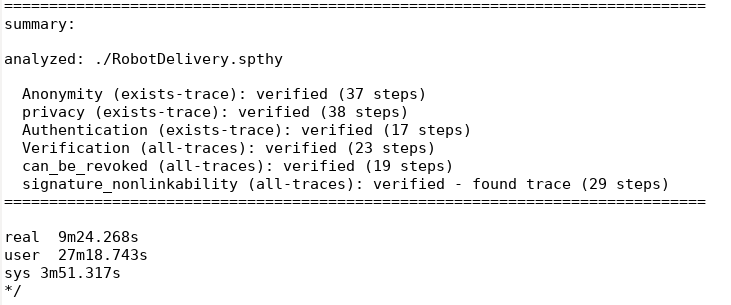}
    \caption{Tamarin Prover Analysis Summary}
    \label{fig:tamarin}
\end{figure*}
Fig. \ref{fig:tamarin} shows outcome summary of the \textit{"RobotDelivery.spthy"} file run over the \textit{tamrin prover} installed in Windows Subsystem for Linux (WSL). To read and get more details, readers can refer to the \cite{Chintan2022} file. An outcome shows that the proposed protocol is verified for all the six lemmas defined for the desired property verification.   
\subsection{Security Property Analysis}
In this subsection, we analyze the proposed protocol using various security properties necessary for any robot-based delivery system. \begin{itemize}
\item \textit{Customer Anonymity and Privacy:}
In the proposed scheme, we don't share any identity-related information over the public channel in plaintext. The DID is registered in and resolved through the private blockchain. The customer purchase details are also verified in the hash format; hence there is no chance for any insider to see the customer order details in the plain text. Hence customer anonymity and customer privacy are achieved.   
\item \textit{Order Unlinkability:}
In the proposed scheme, for any two consecutive orders $O_1$ and $O_2$, the $DID_u^{O_1}$ $\neq$  $DID_u^{O_2}$. For each new order, a new DID is generated. Thus there is no chance of identity linking. In each order $O_i$, we take $SIP_h$($PD_i$) to verify the product details, so the hash's randomness leads to the product's unlinkability. 
\item \textit{Secure Customer Authentication and Verification:}
In the proposed scheme, the customer presents a verifiable credential to the $VC_u$ CSP during the order placing phase, and the CSP authenticates it through the full proof DID resolution technique. During the customer verification phase, the customer presents the $OSVC$ signed by CSP to the robot, and those $OSVC$ are also verified through the DID resolution method; hence the customer is authenticated two times in the complete product purchase life cycle. 
\item \textit{Secure Robot Verification: } For the robot verification, we use the challenge-response-based PUF technique. In the proposed scheme, the CSP selects random CRP from the large CRP database and stores signed challenges into CMA. The CMA uses it during product delivery. The complete robot verification occurs through the CSP, and there is no direct interaction between the robot with the customer. Hence, the proposed scheme achieves secure robot verification. 
\item \textit{Non-Repudiation: } In the proposed scheme, both verifiable credentials ($VC_u$ and $OSVC$) are signed by the private key of the CSP, so there is no option available with the CSP to deny signing those credentials. Similarly, the user signs random nonce $n_u$ as a challenge using their private key, so there is no option available to the customer also of denying the signature. Hence, the proposed scheme achieves non-repudiation or non-deniability. 
\item \textit{Secure against forgery attacks: } In the proposed scheme, both the verifiable credentials ($VC_u$ and $OSVC$) are issued by the CSP, and its private key signs them both. Hence, no option is available to any customer to forge/clone the CSP generated verifiable credentials.  
\item \textit{Secure against VCs loss/stolen:} The $VC$ and the $OSVC$ are stored in the CMA, and CMA is available only to the customer. At the same time, $DID_u$ is only available to the customer and its new for each order. So, even if an attacker captures any of $VC$ or $OSVC$, it is impossible to prove them to be a legitimate user. In the proposed scheme, we verify product details as well. So, it is near impossible for the attacker to get the credentials and use them without full access to CMA. 
\item \textit{Credential Revocation : } In the proposed scheme, both verifiable credentials ($VC_u$ and $OSVC$) issued to the customer are signed by the $Pr_{R_{csp}}$ and $Pr_{R_{csp}}$ is generated every time new for each customer order. And, the CSP can any time terminate the entry of its $DID_{csp}$ and $Pr_{V_{csp}}$ from the blockchain through the blockchain administrator. Hence after termination, the customer can not use $VC_u$ and $OSVC$ for further verification.  
\end{itemize} 

The proposed scheme also achieves security against impersonation attack (only the user knows $(VC_u)$ and $OSVC$, forgery attack, reply attack (use of $T_S$ and $n_u$), accountability (using ($VC_u)^{pr_{Gov}}$ verification by CSP), minimal disclosure (only limited information disclosed by the customer), and security against reuse of credentials (for each order new $VC_u$ and $OSVC$).       

\section{Experimental Study and Performance Analysis}
\noindent This section demonstrates the experimental work related to proposed user verification and the proposed robot verification approach. We first major the computation cost of the proposed work based on several cryptographic operations. Then, we major the computation time required by our implemented work using the $big_O$ python module. Following it, we have also analyzed the communication cost in terms of the number of bits communicated over public channels. There is an option of accuracy analysis in robot verification and user verification using machine learning, but that we have considered as future work and not provided in this work.  
\subsection{Experimental Setup}
As a mobile customer agent, we used the Blockstream Green application, which supports multi-signature and two-factor authentication. We have run this application over the android operating system on a mobile device with 12 GB RAM. As a robot, We have used raspberry pi 4 model B with 4 GB RAM that comes with onboard wireless networking and consumes less power than the computer. We have used a DELL Alienware laptop with 12th Generation Intel Core i9-12900HK (24 MB cache, 14 cores, 20 threads, up to 5.00 GHz Turbo) with 64 GB RAM system as a company's service provider server and that was connected with AWS Cloud running a permissioned blockchain setup for implementation of proposed work. For communication between customer, robot, and CSP, we have used publish/subscribe based Message Queuing Telemetry Transport (MQTT) protocol handling communication channels through AWS MQTT broker running on AWS cloud.   
\subsection{Performance Analysis}
We have considered four major parameters for the performance analysis of the proposed work. The first parameter is the end-to-end network latency that involves round trip time required in the order placing phase by the customer, customer verification by the robot, and robot verification by the customer. We have considered the second parameter as a computation cost in each step based on the individual time taken by each cryptographic operation. The third parameter is communication cost, which presents the total number of bits communicated over the network in all three major phases. And the last parameter is throughput, in which we have analyzed the impact on end-to-end latency over the growing number of orders.   

\begin{table}[H]
    \centering
     \caption{Individual Computation Cost}
    \begin{tabular}{|c|c|c|c|} \hline
    \textbf{Operation} & \textbf{Customer} & \textbf{CSP} & \textbf{Robot} \\ \hline
    Hash Operation  & 1.01 ms & 0.05 ms & - \\ \hline
    ECDSA Sign  & 8.19 ms & 2.03 ms & - \\ \hline
    ECDSA Verify  & 11.61 ms & 4.12 ms  & 5.23 ms \\ \hline
    ECC Encryption  & 8.67 ms & 6.32 ms & 10.12 ms \\ \hline
    ECC Decryption  & 9.33 ms & 7.47 ms  & - \\ \hline
    \end{tabular}
    \label{tab:comptime}
\end{table}
\noindent The ECDSA verification can take more time for the first time because it also involves DID resolution. The robot takes less time in ECDSA verification because the CSP stores the public key to verify the signature of the CSP. Table \ref{tab:comptime} shows average time required for each operation by all entities and Table \ref{tab:compcostphasewise} represents phase-wise computation cost for the proposed work.  
\begin{table}[H]
    \centering
     \caption{Phase wise Computation Cost}
    \begin{tabular}{|c|c|c|c|} \hline
    \textbf{Phases} & \textbf{Customer} & \textbf{CSP} & \textbf{Robot} \\ \hline
    Order Placing Phase  & 51.43 ms & 18.72 ms & - \\ \hline
    Customer Verification Phase  & 8.67 ms & 17.79 ms & 5.23 ms \\ \hline
    Robot Verification Phase & 57.6 ms & 37.68 ms & 15.35 ms \\ \hline
    \end{tabular}
    \label{tab:compcostphasewise}
\end{table}
We have used key-based SIPHash as a hash function that provides the 64-bit output and leads to 64-bit DID, 64-bit VC, and 64-bit hash output. The size of the retrieved product ID ($PID$) is 64-bit, the size of the challenge and response is 128 bits each, and the size of the random numbers ($n_x$) generated is 64-bit. The verification bit, success or failure signal, and robot verification request are of size 1 bit each. Based on these individual costs, Table \ref{tab:commcost} represents the phase-wise communication cost for the proposed scheme. 
\begin{table}[H]
    \centering
    \caption{Communication Cost}
    \begin{tabular}{|c|c|} \hline
    \textbf{Phases} & \textbf{Communication Cost} \\ \hline
    Order Placing Phase   & 448 bits \\ \hline
    Customer Verification Phase  &  385 bits \\  \hline
    Robot Verification Phase   & 578 bits \\ \hline
    \end{tabular}
    \label{tab:commcost}
\end{table}
For communication purposes, we have used the publish/subscribe based MQTT protocol as an application layer protocol, and the packets were transmitted in the javascript format. The MQTT uses IP and TCP protocols as a network layer and transport layer protocol. All the phases (including the verifiable credential obtaining phase) were tested over the public channel. The end-to-end latency does not include computation time. Following Table \ref{tab:endtoend} presents cumulative end-to-end communication delay (includes all four major networking delays: queuing, processing, transmission and propagation delay) for each phase.  
\begin{table}[H]
    \centering
    \caption{End - to - End Latency}
    \begin{tabular}{|c|c|} \hline
    \textbf{Phases} & \textbf{Communication Cost} \\ \hline
    Customer VC Obtaining Phase & 59 ms \\ \hline
    Order Placing Phase   & 79 ms \\ \hline
    Customer Verification Phase  &  102 ms \\  \hline
    Robot Verification Phase   & 93 ms \\ \hline
    \end{tabular}
    \label{tab:endtoend}
\end{table}
We have considered throughput as a distribution variation in the end-to-end latency for each unit of increasing orders.  
\begin{figure}
  \centering
  \includegraphics[width=\linewidth]{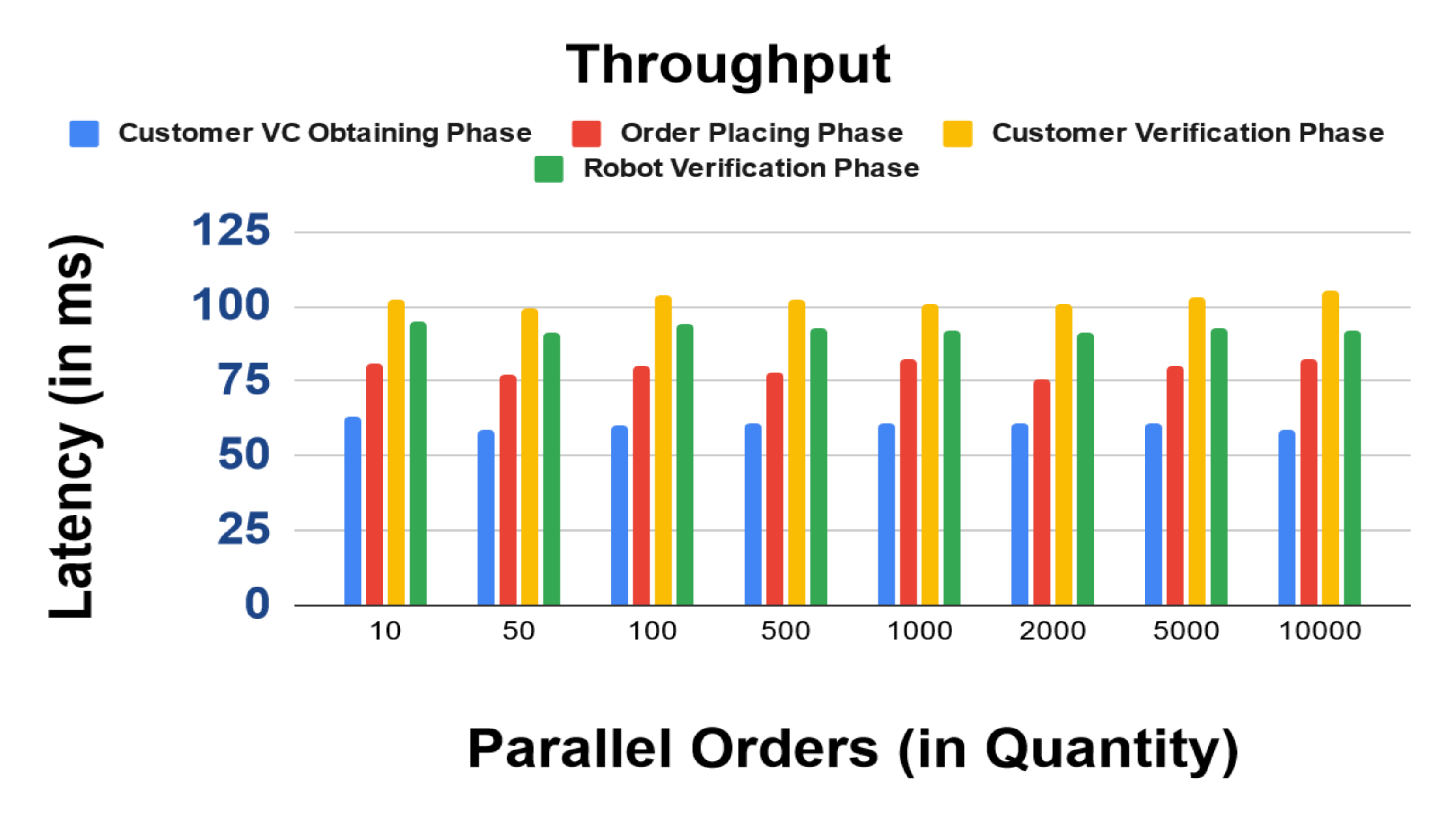}
 \caption{Throughput}
 \label{Fig:throughput}
\end{figure}
Fig. \ref{Fig:throughput} presents the variation in end-to-end latency for each phase over the increasing number of orders ranging from ten requests to ten-thousand requests in parallel. It shows a minor variation in latency for any number of orders in each phase.    
\section{Conclusion and Open Challenges}
\noindent In this section, we present a cursory discussion about the contribution of this work and open challenges related to this work. The open challenges can be much more, highlighting the foremost ones related to contributed work. 
\subsection{Conclusions}
The robot-based delivery system ensures cost-effective, contactless, on-time delivery for e-commerce companies. In this manuscript, we proposed a first-time DID-based customer verification with PUF based robot verification system. We proposed a novel DID-based framework and a DID-based scheme for secure product delivery after authenticated customer verification. We also presented security verification using Tamarin prover and key security property analysis. We discussed the performance analysis of the proposed scheme based on end-to-end latency, computation cost, communication cost, and throughput and proceeded with the implementation of the proposed work using real-time hardware and python language. Overall, this is the first and novel DID-based solution for secure robot-based product delivery. There are some open issues related to the work that we presented in the following subsection. 

\subsection{Open Challenges and extension possibilities}
In the proposed work, 
\begin{enumerate}
\item we have contemplated that the customer can not cancel the order once the robot leaves for delivery. In the future, we will provide a solution so that customers can cancel/update their orders anytime. In that solution, We will provide a novel cryptographic dynamic accumulator for the e-commerce company to revoke the customer in such a way that the customer verification will not succeed if the customer cancels/updates the order even after the robot starts its product delivery journey. 
\item We have considered the service provider entity a VCs issuer and verifier. There is nothing wrong because issuers of VCs are still trusted entities in DID mechanism. In the future, we will introduce a separate entity for issuing VCs and verifying VCs. 
\item We haven't discussed about loss of private key by customer. If customer loss his/her private key then it is difficult for the customer to recover the private key and prove himself/herself as a legitimate customer. For that, we will introduce a biometric DID as a tool to recover the lost key with out any extra backup storage.  
\item We have not discussed any work related to sharing OSVC by the verified customer to a legitimate receiver (for example, a friend or family member) in the customer's dearth. So in the future, we will also introduce a solution for that challenge. 
\item We haven't discussed anything related to DID's auditability and privacy issues. If DID of the customer is stolen, it can become difficult to detect misuse of IT by an adversary to communicate with CSP. 
\item We haven't provided a comparative performance analysis of the proposed work because, as per our knowledge, this is the first decentralized identifier paper for customer and robot verification.  
\item We haven't used zk-SNARK or Zero-Knowledge Proof (ZKP) for verification of credentials, and it is an interesting dimension to use ZKP for secured anonymous verification of verifiable credentials. 
\item We haven't explored the key sharing aspect followed by symmetric encryption. So, the researcher can explore the symmetric encryption after authenticated key sharing for the DID-based customer verification in e-commerce product delivery.    
\item The CSP stores its public key in the robot. There is a high chance that if an attacker gets access to the robot (ideally, it is near impossible), he may delete the public key. In the proposed scheme, we have considered that if any such incident occurs, the robot will ask the CSP over the public channel for the public key. We can also come up with any other better solution as future work to this problem.   
\item We have presented a scenario of the robot-based delivery system, but the proposed work can be extended to the drone-based delivery system, and it can also come with numerous novel open challenges. To the best of our knowledge, there does not exist any DID based drone verification or user verification schemes.  
\end{enumerate}

\bibliographystyle{IEEEtran}
\bibliography{sample-base}
\end{document}